\documentclass[aps,prl,twocolumn,showpacs,amsmath,amssymb]{revtex4-1}
\usepackage{amsmath}
\usepackage{graphicx}
\usepackage{subfigure}
\usepackage{epstopdf}
\usepackage{color}
\usepackage{multirow}
\usepackage{setspace}
\usepackage{overpic}
\usepackage{amssymb}
\usepackage{lineno}
\usepackage{bm}
\usepackage{rotating}
\usepackage[utf8]{inputenc}
\let\oldequation\equation
\let\oldendequation\endequation

\renewenvironment{equation}
  {\linenomathNonumbers\oldequation}
  {\oldendequation\endlinenomath}

\begin{document}

\title{\bf \boldmath
First Observation of $D^+\to \eta\mu^+\nu_\mu$ and Measurement of its Decay Dynamics}

\author{
M.~Ablikim$^{1}$, M.~N.~Achasov$^{10,c}$, P.~Adlarson$^{64}$, S. ~Ahmed$^{15}$, M.~Albrecht$^{4}$, A.~Amoroso$^{63A,63C}$, Q.~An$^{60,48}$, ~Anita$^{21}$, X.~H.~Bai$^{54}$, Y.~Bai$^{47}$
% ...
\\
\vspace{0.2cm}
(BESIII Collaboration)\\
\vspace{0.2cm} {\it
$^{1}$ Institute of High Energy Physics, Beijing 100049, People's Republic of China\\
% ...
$^{4}$ Bochum Ruhr-University, D-44780 Bochum, Germany\\
% ...
$^{10}$ G.I. Budker Institute of Nuclear Physics SB RAS (BINP), Novosibirsk 630090, Russia\\
% ...
$^{15}$ Helmholtz Institute Mainz, Johann-Joachim-Becher-Weg 45, D-55099 Mainz, Germany\\
% ...
$^{21}$ Indian Institute of Technology Madras, Chennai 600036, India\\
% ...
$^{47}$ Southeast University, Nanjing 211100, People's Republic of China\\
$^{48}$ State Key Laboratory of Particle Detection and Electronics, Beijing 100049, Hefei 230026, People's Republic of China\\
% ...
$^{54}$ University of Jinan, Jinan 250022, People's Republic of China\\
% ...
$^{60}$ University of Science and Technology of China, Hefei 230026, People's Republic of China\\
% ...
$^{63}$ (A)University of Turin, I-10125, Turin, Italy; (B)University of Eastern Piedmont, I-15121, Alessandria, Italy; (C)INFN, I-10125, Turin, Italy\\
$^{64}$ Uppsala University, Box 516, SE-75120 Uppsala, Sweden\\
% ...
$^{c}$ Also at the Novosibirsk State University, Novosibirsk, 630090, Russia\\
% ...
}
}

\begin{abstract}
  By analyzing a data sample corresponding to an integrated luminosity
  of $2.93~\mathrm{fb}^{-1}$ collected at a center-of-mass energy of
  3.773 GeV with the BESIII detector, we measure for the first time
  the absolute branching fraction of the $D^+\to \eta \mu^+\nu_\mu$
  decay to be ${\mathcal B}_{D^+\to \eta
    \mu^+\nu_\mu}=(10.4\pm1.0_{\rm stat}\pm0.5_{\rm syst})\times
  10^{-4}$.  Using the world averaged value of ${\mathcal B}_{D^+\to
    \eta e^+\nu_e}$, the ratio of the two branching fractions is
  determined to be ${\mathcal B}_{D^+\to \eta \mu^+\nu_\mu}/{\mathcal
    B}_{D^+\to \eta e^+\nu_e}=0.91\pm0.13_{\rm (stat+syst)}$, which
  agrees with the theoretical expectation of lepton flavor
  universality within uncertainty.
  By studying the differential decay
  rates in five four-momentum transfer intervals, we obtain the
  product of the hadronic form factor $f^{\eta}_{+}(0)$ and the $c\to
  d$ Cabibbo-Kobayashi-Maskawa matrix element $|V_{cd}|$ to be
  $f_{+}^\eta (0)|V_{cd}|=0.087\pm0.008_{\rm stat}\pm0.002_{\rm
    syst}$. Taking the input of $|V_{cd}|$ from the global fit in the
  standard model, we determine $f_{+}^\eta (0)=0.39\pm0.04_{\rm
    stat}\pm0.01_{\rm syst}$. On the other hand, using the value of
  $f_+^{\eta}(0)$ calculated in theory, we find
  $|V_{cd}|=0.242\pm0.022_{\rm stat}\pm0.006_{\rm syst}\pm0.033_{\rm
    theory}$.
\end{abstract}

\pacs{13.20.Fc, 12.15.Hh}

\maketitle

\oddsidemargin  -0.2cm
\evensidemargin -0.2cm

In the standard model (SM), the couplings between three families of
leptons and gauge bosons are independent of lepton flavors. This
property is known as lepton flavor universality
(LFU)~\cite{Salam1964,Fajfer2012,Fajfer2015,Guo2017}. Semileptonic
(SL) decays of pseudoscalar mesons, which are well understood in the
SM, offer an ideal platform to test LFU and search for new physics
effects. In the past decade, the BaBar, Belle, and LHCb collaborations
reported anomalies in LFU tests with various SL $B$ decays. The
measured branching fraction (BF) ratios ${\mathcal
  R}_{\tau/\ell}={\mathcal B}_{B\to \bar
  D^{(*)}\tau^+\nu_\tau}/{\mathcal B}_{B\to \bar
  D^{(*)}\ell^+\nu_\ell}$~($\ell=\mu$,
$e$)~\cite{babar_1,babar_2,lhcb_1,belle2015,belle2016,lhcb_1a,belle2019}
deviate from the SM predictions by $3.1\sigma$~\cite{hflav2018}.
Various
models~\cite{BFajfer2012,Fajfer2012,Celis2013,Crivellin2015,Crivellin2016,Bauer2016}
were proposed to explain these differences.
In view of this, scrutinizing the ratios of the semimuonic $D$ decay BFs over their corresponding semielectronic counterparts offers important complementary tests of $e{\text -}\mu$ LFU.
Recently, BESIII
reported tests of LFU with the SL decays $D\to \bar K
\ell^+\nu_\ell$~\cite{epjc76,bes3-D0-Kmuv} and $D\to \pi
\ell^+\nu_\ell$~\cite{bes3-pimuv}.
Using the world averaged BFs~\cite{pdg2018}, the difference between the BF ratio of $D\to \pi\ell^+\nu_\ell$ decays which are mediated via $c\to d\ell^+\nu_\ell$ (${\mathcal R}_{\mu/e}^{c\to d}={\mathcal B}_{D\to \pi\mu^+\nu_\mu}/{\mathcal B}_{D\to \pi e^+\nu_e}$) and the SM prediction is below $2\sigma$.
Meanwhile, there is still no experimental confirmation of the $D^+\to\eta\mu^+\nu_\mu$ decay,
although it was predicted in the quark model 30 years ago~\cite{isgw}.
Verification of  ${\mathcal R}_{\mu/e}^{c\to d}$ with $D^+\to\eta\ell^+\nu_\ell$ decays,  which is also mediated via $c\to d\ell^+\nu_\ell$, is key to clarifying these situations.
 In this Letter, we report a complementary test
of LFU with $D^+\to\eta \ell^+\nu_\ell$ decays based on the first
measurement of the BF of $D^+\to\eta \mu^+\nu_\mu$. Throughout this
Letter, charge conjugate channels are always implied.  The BF obtained
will also be important for the determination of the $\eta$-$\eta'$
mixing angle, which will benefit the understanding of nonperturbative
quantum-chromodynamics~(QCD) effects~\cite{cancel}.

The investigation of $D^+\to\eta \mu^+\nu_\mu$ decay dynamics allows
the determination of the $c\to d$ Cabibbo-Kobayashi-Maskawa (CKM)
matrix element $|V_{cd}|$ and the hadronic form factor (FF)
$f^{\eta}_{+}(0)$. The value of $f^{\eta}_{+}(0)$ has been calculated
with various approaches, e.g., QCD light-cone sum rules
(LCSR)~\cite{ylwu,dse:2015gdu,dse:2013nof}, light-front quark model
(LFQM)~\cite{dse:2012rcv}, covariant confined quark model
(CCQM)~\cite{dse:2018nrs,dse:2018nrsnew}, and relativistic quark model~(RQM)~\cite{rqm}. The predicted values vary
in a wide range from 0.36 to 0.71.
According to Refs.~\cite{Koponen1,Koponen2}, the predicted FFs of the
SL $D$ decays are expected to be insensitive to the spectator quark.
Measurement of the hadronic FF in $D^+\to\eta \mu^+\nu_\mu$ decay can be used to distinguish between these calculations.  The predicted FF that passes the experimental test is useful to determine $|V_{cd}|$.
Conversely, measurements of SL $D$ decay hadronic FFs help constrain lattice QCD
calculations and lead to more reliable calculations of the hadronic
FFs of SL $D$ and $B$ decays, which are crucial to accurately determine
CKM parameters~\cite{Koponen1,Koponen2,Brambilla,Bailey},
and test the unitarity of the CKM matrix.

In this analysis, we use a data sample corresponding to an integrated
luminosity of 2.93~fb$^{-1}$~\cite{lum_bes3} taken at the
center-of-mass energy $\sqrt s=3.773$~GeV with the BESIII detector.
Details about the design and performance of the BESIII detector are
given in Ref.~\cite{BESCol}.  Simulated samples produced with the {\sc
  geant4}-based~\cite{geant4} Monte Carlo (MC) package which includes
the geometric description of the BESIII detector and the detector
response, are used to determine the detection efficiency and to
estimate the backgrounds. The simulation includes the beam energy
spread and initial state radiation (ISR) in the $e^+e^-$ annihilations
modeled with the generator {\sc kkmc}~\cite{kkmc}.  The inclusive MC
samples consist of the production of $D^0\bar D^0$, $D^+D^-$, and
non-$D\bar{D}$ decays of the $\psi(3770)$, the ISR production of the
$J/\psi$ and $\psi(3686)$ states, and the continuum processes
incorporated in {\sc kkmc}~\cite{kkmc}.  The known decay modes are
modeled with {\sc evtgen}~\cite{evtgen} using BFs taken from the
Particle Data Group~\cite{pdg2018}, and the remaining unknown decays
from the charmonium states with {\sc lundcharm}~\cite{lundcharm}. The
final state radiation from charged particles is incorporated with the
{\sc photos} package~\cite{photos}.  The vector hadronic FF of the SL
decay $D^+\to \eta\mu^+\nu_\mu$ is simulated with the modified-pole
model~\cite{MPM}, where the parameter $\alpha$ of the vector hadronic
FF is set to be that of $D^+\to \pi^0e^+\nu_e$ measured by
BESIII~\cite{bes3-Dp-piev}, and the pole mass is set at the nominal
$D^{*+}$ mass~\cite{pdg2018}.

The analysis is performed with the double-tag (DT) method, benefiting
from the advantage of $D^+D^-$ pair production at $\sqrt s=3.773$~GeV.
If a $D^-$ meson [called single-tag (ST) $D^-$ meson] is fully
reconstructed by the hadronic decays $D^-\to K^{+}\pi^{-}\pi^{-}$,
$K^0_{S}\pi^{-}$, $K^{+}\pi^{-}\pi^{-}\pi^{0}$,
$K^0_{S}\pi^{-}\pi^{0}$, $K^0_{S}\pi^{+}\pi^{-}\pi^{-}$, and
$K^{+}K^{-}\pi^{-}$, the presence of a $D^+$ meson is guaranteed.  If
the $D^+\to\eta\mu^+\nu_\mu$ decay can be found in the system
recoiling against an ST $D^-$ meson, the candidate event is called a
DT event. The BF of the SL decay is determined by
\begin{equation}
\label{eq:bf}
{\mathcal B}_{D^+\to\eta\mu^+\nu_\mu}=N_{\mathrm{DT}}/(N_{\mathrm{ST}}^{\rm tot}\cdot\varepsilon_{\rm SL} \cdot{\mathcal B}_{\eta\to\gamma\gamma}),
\end{equation}
where $N_{\rm ST}^{\rm tot}$ and $N_{\rm DT}$ are the yields of the ST
and DT candidates in data, respectively, and ${\mathcal
  B}_{\eta\to\gamma\gamma}$ is the BF of the $\eta\to\gamma\gamma$
decay. The
$\varepsilon_{\rm SL}=\Sigma_i[(\varepsilon^i_{\rm DT}\cdot N^i_{\rm ST})/(\varepsilon^i_{\rm
  ST}\cdot N_{\rm ST}^{\rm tot})]$ is the effective signal efficiency of finding
$D^+\to\eta\mu^+\nu_\mu$ decay in the presence of the ST $D^-$ meson.
Here $i$ denotes the tag mode, and $\varepsilon_{\rm ST}$ and $\varepsilon_{\rm DT}$ are the
efficiencies of selecting the ST and DT candidates, respectively.

\begin{figure}[htbp]\centering
\includegraphics[width=1.0\linewidth]{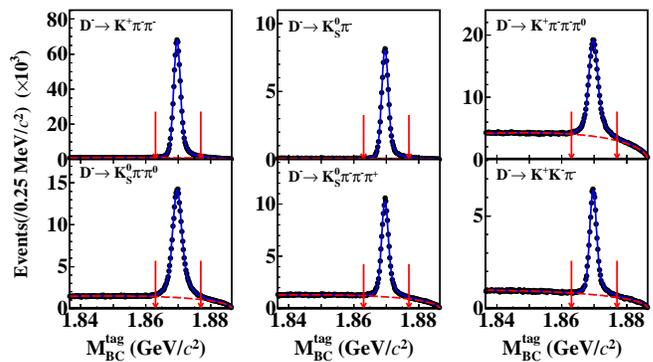}
\caption{ Fits to the $M_{\rm BC}$ distributions of the ST $D^-$
  candidates.  Data are shown as dots with error bars.  The blue solid and
  red dashed curves are the fit results and the fitted backgrounds,
  respectively.  Pairs of red arrows show the $M_{\rm BC}$ windows.
}\label{fig:datafit_Massbc}
\end{figure}

This analysis uses the same $K^\pm$, $\pi^\pm$, $K^0_S$, $\gamma$, and
$\pi^0$ selection criteria as those employed in
Refs.~\cite{epjc76,bes3-pimuv,cpc40,bes3-Dp-K1ev,bes3-etaetapi,bes3-omegamuv}. The
ST $D^-$ mesons are distinguished from combinatorial backgrounds by
using the energy difference $\Delta E\equiv
E_{D^-}-E_{\mathrm{beam}}$ and the beam-constrained mass $M_{\rm
  BC}\equiv\sqrt{E_{\mathrm{beam}}^{2}-|\vec{p}_{D^-}|^{2}}$. Here,
$E_{\mathrm{beam}}$ is the beam energy, and $E_{D^-}$ and
$\vec{p}_{D^-}$ are the total energy and momentum of the ST $D^-$
candidate in the $e^+e^-$ center-of-mass frame. If multiple
combinations for an ST mode are present in an event, the combination
with the smallest $|\Delta E|$ per tag mode per charge is retained for
further analysis. The candidates are required to satisfy $\Delta E\in
(-0.055,0.045)$~GeV for the tags containing $\pi^0$ and $\Delta E\in
(-0.025,0.025)$~GeV for the other tags. For each tag mode, the yield
of ST $D^-$ mesons is determined from the maximum likelihood fit of
the $M_{\rm BC}$ distribution of the accepted candidates. In the fit,
the signal and background are described by  an MC-simulated shape and
an ARGUS function~\cite{argus}, respectively. To take into account the
resolution difference between data and MC simulation, the MC-simulated
signal shape is convolved with a double-Gaussian function.
The widths and relative abundances of the Gaussian components are
free parameters of the fit.
The resulting fits of these $M_{\rm BC}$ distributions are exhibited in
Fig.~\ref{fig:datafit_Massbc}. The candidates with $M_{\rm
  BC}\in(1.863,1.877)$ GeV/$c^2$ are kept for further analysis. The
total yield of ST $D^-$ mesons is $N^{\rm tot}_{\rm
  ST}=1522474\pm2125_{\rm stat}$.

At $\sqrt s=3.773$~GeV, the $D^+$ and $D^-$ mesons are in back-to-back direction. The $D^+\to \eta\mu^+\nu_\mu$ candidates are selected in the sides
recoiling against the ST $D^-$ mesons.  It is required that there is
only one charged track available for muon identification.  The muon
candidate is required to satisfy $|V_{xy}|<1$~cm and $|V_{z}|<10$~cm,
where $|V_{xy}|$ and $|V_{z}|$ are the distances of closest approach
to the interaction point of the reconstructed track in the transverse
plane and along the axis of the drift chamber, respectively. Its polar angle
($\theta$) with respect to the axis of the drift chamber must be within
$|\cos\theta|<0.93$.

Muon identification uses information from the time-of-flight and the
electromagnetic calorimeter~(EMC), as well as the specific ionization
energy loss measured in the main drift chamber. The combined
confidence levels for various particle hypotheses~($CL_i$, $i=e$,
$\mu$, and $K$) are calculated. Muon candidates are required to
satisfy $CL_{\mu}>$ 0.001, $CL_{\mu}>CL_e$, and $CL_{\mu}>CL_K$.
To reduce misidentification between hadrons and muons, the deposited energy in the EMC of the muon candidate is required to be within
(0.105,\,0.275)\,GeV, since it is expected to concentrate around 0.2 GeV.

The $\eta$ candidates are reconstructed via the $\eta\to\gamma\gamma$
decay.  The invariant mass of the $\gamma\gamma$ candidate is required
to be within $(0.510,\,0.570)$\,GeV$/c^{2}$. To improve momentum
resolution, a one-constraint~(1-C) kinematic fit is done on the
selected photon pair, whose invariant mass is constrained to the
$\eta$ nominal mass ($m_\eta$)~\cite{pdg2018}.

Due to the misidentification of pions as muons, some hadronic $D^+$ decays survive the above selection criteria.
To suppress the peaking backgrounds from $D^+\to\eta\pi^+$ decays,
we require the $\eta\mu^+$ invariant mass~($M_{\eta\mu^+}$) to be less than 1.74 GeV/$c^2$.
To reject the backgrounds containing $\pi^{0}$, e.g., $D^+\to\eta\pi^+\pi^0$, we require that
the maximum energy of any extra photon~($E_{\rm extra~\gamma}^{\rm max}$) is less than 0.30~GeV and
there is no extra $\pi^0$ ($N_{\rm extra~\pi^0}$) in the candidate event.
Here, the extra photon and $\pi^0$ denote the ones which have not been used in the DT selection.

The number of SL decays is determined using a kinematic quantity
defined as $U_{\mathrm{miss}}\equiv
E_{\mathrm{miss}}-|\vec{p}_{\mathrm{miss}}|$, which is expected to
peak around 0 for the correctly reconstructed signal
events. Here, $E_{\mathrm{miss}}\equiv
E_{\mathrm{beam}}-E_{\eta}-E_{\mu^{+}}$ and
$\vec{p}_{\mathrm{miss}}\equiv
\vec{p}_{D^+}-\vec{p}_{\eta}-\vec{p}_{\mu^{+}}$ are the missing energy
and momentum of the DT event in the $e^+e^-$ center-of-mass frame, in
which $E_{\eta\,(\mu^+)}$ and $\vec{p}_{\eta\,(\mu^+)}$ are the energy
and momentum of the $\eta$\,($\mu$) candidates. The
$U_{\mathrm{miss}}$ resolution is improved by constraining the $D^+$
energy to the beam energy and $\vec{p}_{D^+} \equiv
{-\hat{p}_{D^-}}\cdot\sqrt{E_{\mathrm{beam}}^{2}-m_{D^+}^{2}}$, where
$\hat{p}_{D^-}$ is the unit vector in the momentum direction of the ST
$D^-$ and $m_{D^+}$ is the $D^+$ nominal mass~\cite{pdg2018}.

\begin{figure}[htbp] \centering
\includegraphics[width=1.0\linewidth]{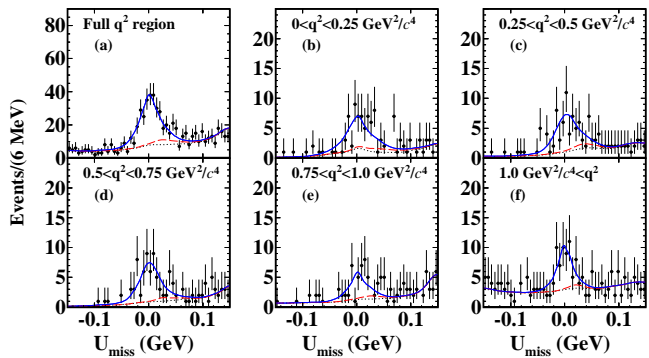}
\caption{Fits to the $U_{\rm miss}$ distributions of the
  $D^{+}\rightarrow\eta\mu^{+}\nu_{\mu}$ candidate events.  Data are
  shown as dots with error bars.  In each figure, the blue solid curve
  is the fit result, the black dotted curve is the fitted
  non-peaking background, and the difference between red dashed and
  black dotted curves is the peaking background of $D^+\to
  \eta\pi^+\pi^0$.}
\label{fig:fit_Umistry1}
\end{figure}

Figure~\ref{fig:fit_Umistry1}\,(a) shows the $U_{\mathrm{miss}}$
distribution of the accepted DT events in data. The SL decay yield is
obtained from an unbinned fit to the $U_{\mathrm{miss}}$
distribution, where the SL signal, peaking backgrounds of $D^+\to
\eta\pi^{+}\pi^{0}$, and non-peaking backgrounds (including a small
contribution from wrongly reconstructed ST candidates) are described by
the corresponding MC-simulated shapes.  The yields of the signal and
non-peaking backgrounds are free parameters of the fit,
while the yield of the peaking
background from $D^+\to \eta\pi^{+}\pi^{0}$ decays is fixed based on
MC simulation. The fit result is shown in
Fig.~\ref{fig:fit_Umistry1}(a). From the fit, we obtain the yield of
DT events $N_{\rm DT}=234\pm22_{\rm stat}$. The statistical
significance, calculated by $\sqrt{-2{\rm ln ({\mathcal L_0}/{\mathcal
      L_{\rm max}}})}$, is found to be greater than $10\sigma$. Here,
${\mathcal L}_{\rm max}$ and ${\mathcal L}_0$ are the maximal
likelihood of the nominal fit and that of the fit without signal
component, respectively.

Table 1 of Supplemental Material \cite{sup} shows tag dependent numbers of $N^i_{\rm ST}$,
$\epsilon^i_{\rm ST}$, and $\epsilon^i_{\rm DT}$.
The average efficiency of detecting $D^+\to\eta\mu^+\nu_\mu$ decays is
$\varepsilon_{\rm SL}=0.3752\pm0.0013$.
Here, the efficiency does not
include the BF of the $\eta\to\gamma\gamma$ decay.  To verify the
reliability of the efficiency determination, we have compared
distributions of momenta and $\cos\theta$ of the $\eta$ and $\mu^+$ of
the selected $D^+\to\eta\mu^+\nu_\mu$ candidate events between data
and MC simulation, and they are in good agreement.

Inserting $N_{\rm DT}$, $\varepsilon_{\rm SL}$, $N_{\rm ST}^{\rm tot}$, and the world average of ${\mathcal B}_{\eta\to\gamma\gamma}=0.3941\pm0.0020$~\cite{pdg2018} into Eq.~(\ref{eq:bf}), we obtain
\begin{equation}
{\mathcal B}_{D^+\to\eta \mu^+\nu_\mu}=(10.4\pm1.0_{\rm stat}\pm0.5_{\rm syst})\times 10^{-4}.\nonumber
\end{equation}
In the BF measurement, the systematic uncertainties arise from the
following sources. The uncertainty in the total yield of ST $D^-$
mesons has been studied in Refs.~\cite{epjc76,cpc40,bes3-pimuv}, and
is assigned as 0.5\%. The muon tracking (PID) efficiencies are studied
by analyzing $e^+e^-\to\gamma\mu^+\mu^-$ events, and the muon
tracking (PID) efficiency uncertainty is taken as 0.2\% (0.2\%) per
muon, where the data/MC differences of the two-dimensional (momentum
and $\cos\theta$) distributions of the control samples have been
re-weighted by those of the $D^+\to \eta \mu^+\nu_\mu$ signal decays.
The uncertainty of $\eta$ reconstruction is assumed to be 2.0\%, the
same as $\pi^0$ reconstruction, which was studied with DT $D\bar D$
hadronic decays of $D^0\to K^-\pi^+$, $K^-\pi^+\pi^+\pi^-$ vs. $\bar
D^0\to K^+\pi^-\pi^0$, $K^0_S\pi^0$~\cite{epjc76,cpc40}. The
uncertainties of the requirements of $E_{\rm extra~\gamma}^{\rm max}$
and $N_{\rm extra~\pi^0}$ are estimated to be 2.3\% by analyzing the
DT candidate events of $D^+\to \eta\pi^+$ and $\pi^0 e^+\nu_e$.  The
uncertainty due to the $M_{\eta\mu^+}$ requirement is evaluated by
replacing the nominal requirement with $M_{\eta \mu^+}<1.69~{\rm
  GeV}/c^2$ or $M_{\eta \mu^+}<1.79~{\rm GeV}/c^2$, and the associated
uncertainty is found to be negligible. The uncertainty in the $U_{\rm
  miss}$ fit is assigned to be 3.7\%, which is estimated with
alternative signal and background shapes. For alternative background shapes, the uncertainty due to the peaking background of $D^+\to \eta\pi^+\pi^0$ has been taken into account via varying the quoted BF by $\pm 1\sigma$. The uncertainty due to the
limited MC statistics is 0.5\%. The uncertainty in the MC model, 0.3\%, is
assigned as the difference between our nominal DT efficiency and the
DT efficiency determined by re-weighting the $q^2$ ($q=p_{D^+}-p_{\eta}$ is the total
four momentum of $\mu^+\nu_\mu$) distribution of the signal MC events
using the FF parameters obtained from data.  Adding these
uncertainties quadratically yields the total systematic uncertainty to
be 4.9\%.

To study the dynamics in $D^+\to\eta \mu^+\nu_\mu$ decay, the SL
candidate events are divided into five $q^2$ intervals: $(0.0,0.25)$,
$(0.25,0.5)$, $(0.5,0.75)$, $(0.75,1.0)$, and $(1.0,(m_{D^+}-m_\eta)^2)$~GeV$^2$/$c^4$.
The partial decay
rate in the $i$th $q^2$ interval, $\Delta\Gamma_{\rm measured}^i$, is
determined by
\begin{equation} \label{eq2}
\Delta\Gamma^{i}_{\rm measured}=N_{\mathrm{produced}}^{i}/(\tau_{D^+} \cdot N_{\mathrm{ST}}^{\rm tot}),
\end{equation}
where $N_{\mathrm{produced}}^{i}$ is the $D^+\to \eta\mu^+\nu_\mu$
signal yield produced in the $i$th $q^{2}$ interval in data,
$\tau_{D^+}$ is the lifetime of $D^+$, $N_{\mathrm{ST}}^{\rm tot}$
is the total yield of ST $D^-$ mesons, and
\begin{equation}\label{eq3}
N_{\mathrm{produced}}^{i}=\sum_{j}^{N_{\mathrm{intervals}}}(\varepsilon^{-1})_{ij}N_{\mathrm{observed}}^{j},
\end{equation}
where $N_{\rm observed}^j$ is the $D^+\to \eta\mu^+\nu_\mu$ signal yield observed in the $j$th $q^{2}$ interval
and $\varepsilon$ is the efficiency matrix (Table 2 of Supplemental Material \cite{sup})
given by
\begin{equation} \label{eq4}
\varepsilon_{ij}=\sum_k
\left[(N^{ij}_{\mathrm{reconstructed}} \cdot N_{\rm ST})/(N^{j}_{\mathrm{generated}} \cdot \varepsilon_{\mathrm{ST}})\right]_k/N_{\rm ST}^{\rm tot}.
\end{equation}
Here, $N^{ij}_{\mathrm{reconstructed}}$ is the $D^+\to \eta\mu^+\nu_\mu$ signal yield
generated in the $j$th $q^{2}$ interval and reconstructed in the $i$th $q^{2}$ interval,
$N^{j}_{\mathrm{generated}}$ is the total signal yield generated in the $j$th $q^{2}$ interval, and the index $k$ sums over all ST modes.

\begin{figure}[htbp] \centering
\includegraphics[width=1.0\linewidth]{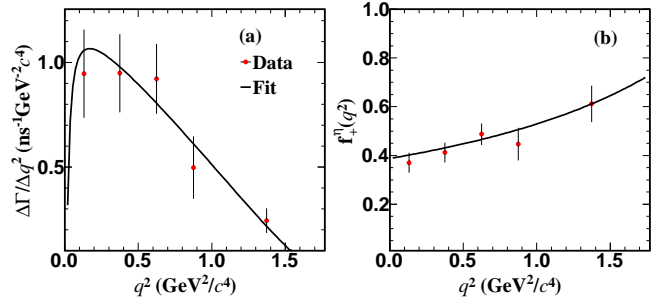}
\caption{ (a) Fit to the partial decay rates and (b) projection to
  $f_+^\eta(q^2)$ for $D^+\to\eta\mu^+\nu_\mu$.  Dots with error bars
  are data, and the solid curves are the fit results.  }
\label{fig:fitdecayrate}
\end{figure}

$N_{\mathrm{observed}}^{i}$ is obtained from the fit to the
$U_{\mathrm{miss}}$ distribution of the
$D^{+}\rightarrow\eta\mu^{+}\nu_{\mu}$ candidate events in the $i$th
$q^{2}$ interval. The fit results of the $U_{\mathrm{miss}}$
distributions in various intervals are shown in
Figs.~\ref{fig:fit_Umistry1}\,(b)-(f), and the partial decay rates
obtained are shown in Fig.~\ref{fig:fitdecayrate}.

With the $\Delta\Gamma^i_{\rm measured}$ obtained above and the partial decay rate $\Delta\Gamma^i_{\rm expected}$ predicted by theory, the $\chi^{2}$ is constructed as
\begin{eqnarray}
\label{eq:chisq}
	\chi^{2}=\sum_{i,j=1}^{5}&(&\Delta\Gamma^{i}_{\mathrm{measured}}-\Delta\Gamma^{i}_{\mathrm{expected}})
	 C_{ij}^{-1} \nonumber \\
&(&\Delta\Gamma^{j}_{\mathrm{measured}}-\Delta\Gamma^{j}_{\mathrm{expected}}),
\end{eqnarray}
\hspace{-0.2cm} where $C_{ij} =
C_{ij}^{\mathrm{stat}}+C_{ij}^{\mathrm{syst}}$ is the covariance
matrix of the measured partial decay rates among $q^2$ intervals, and
\begin{widetext}
\begin{equation}
\Delta\Gamma^{i}_{\mathrm{expected}} = \int_{q_{\mathrm{min}(i)}^{2}}^{q_{\mathrm{max}(i)}^{2}}
\left \{ \frac{G_{F}^{2}|V_{cd}|^{2}}{24\pi^{3}}\cdot \frac{(q^{2}-m^{2}_{\mu})^2\sqrt{E^{2}_{\eta}-m^{2}_{\eta}}}{q^{4}m^{2}_{D}} \cdot
\left [\left (1+\frac{m^{2}_{\mu}}{2q^{2}}\right )m^{2}_{D}(E^{2}_{\eta}-m^{2}_{\eta})|f^\eta_{+}(q^{2})|^{2}\right ] \right \} dq^{2},
\end{equation}
\end{widetext}
where $G_F$ is the Fermi coupling constant; $m_\mu$ is the $\mu^+$ mass; $|\vec p_{\eta}|$ and $E_{\eta}$ are the momentum and energy of $\eta$ in the rest frame of $D^+$, respectively; the vector hadronic FF $f_+^{\eta}(q^2)$ is
formulated following Ref.~\cite{formfactor}.
Here, the scalar hadronic FF $f_0^{\eta}(q^2)$ has been ignored due to negligible sensitivity with limited data.

The FF of $f_+^{\eta}(q^2)$ can be parameterized by the series
expansion~\cite{SEM}, which was widely used in previous analyses and
verified to be consistent with constraints from
QCD~\cite{CLEO-SL,BABR-SL,bes3-D0-piev}. Due to limited data, we adopt
the two-parameter series expansion form
\begin{equation}
f^\eta_{+}(q^2)=\frac{f^\eta_{+}(0)P(0)\Phi(0,t_{0})}{P(q^2)\Phi(q^2,t_{0})}\cdot \frac{1+r_{1}(t_{0})z(q^2,t_{0})}{1+r_{1}(t_{0})z(0,t_{0})},
\end{equation}
where
$t_{0}=t_{+}(1-\sqrt{1-t_{-}/t_{+}})$, $t_{\pm}=(m_{D^+}\pm m_{\eta})^{2}$, and
the functions $P(q^2)$, $\Phi(q^2, t_0)$, and $z(q^2, t_0)$ are defined following Ref.~\cite{SEM}.

The statistical covariance matrix is constructed as
\begin{equation}
C_{ij}^{\rm stat} = (\frac{1}{\tau_{D^{+}} \cdot N_{\mathrm{ST}}^{\rm tot}})^{2}\sum_{n}\varepsilon_{in}^{-1}\varepsilon_{jn}^{-1}(\sigma(N_{\mathrm{obs}}^{n}))^{2},
\end{equation}
where $n$ sums from 1 to 5 intervals.
The systematic covariance matrix is obtained by summing over that of each systematic uncertainty source, which is taken as
\begin{equation}
C_{ij}^{\mathrm{syst}}=\delta(\Delta\Gamma^{i}_{\rm measured})\delta(\Delta\Gamma^{j}_{\rm measured}),
\end{equation}
where $\delta(\Delta\Gamma^{i}_{\rm measured})$ is the systematic uncertainty of the partial decay rate in the $i$th $q^{2}$ interval.
The systematic uncertainties arising from $N_{\rm ST}^{\rm tot}$, $\tau_{D^+}$,  muon tracking and PID, $\eta$ reconstruction, as well as $E_{\rm extra~\gamma}^{\rm max}$ and $N_{\rm extra~\pi^0}$ requirements are taken to be common across all the $q^2$ intervals; while the others are determined separately in each $q^2$ interval as above.

Minimizing the $\chi^2$ in Eq.~(\ref{eq:chisq}) gives the product of
$f_+^\eta (0)|V_{cd}|$ and the first order coefficient $r_1$ to be
$0.087\pm0.008_{\rm stat}\pm0.002_{\rm syst}$ and $-0.9\pm2.7_{\rm
  stat}\pm0.2_{\rm syst}$, respectively. The nominal fit parameters
are taken from the fit with the combined statistical and systematic
covariance matrix, and their statistical uncertainties are taken from
the fit only with the statistical covariance matrix. For each
parameter, the systematic uncertainty is obtained by calculating the
quadratic difference of uncertainties between these two fits.  The fit
result is shown in Fig.~\ref{fig:fitdecayrate} and the goodness-of-fit
is $\chi^2/{\rm NDOF}=1.0/3$, where NDOF is the number of degrees of
freedom.  The $f_+^\eta (0)|V_{cd}|$ measured in this work is
consistent with the measurements using $D^+\to \eta e^+\nu_e$ by
CLEO~\cite{CLEO-etaev} and BESIII~\cite{BESIII-etaev}.

Using our product of $f_{+}^\eta(0)|V_{cd}|$ and the $f_+^{\eta}(0)$ calculated in Ref.~\cite{dse:2018nrsnew}
leads to $|V_{cd}|=0.242\pm0.022_{\rm stat}\pm0.006_{\rm syst}\pm0.033_{\rm theory}$.
This result is consistent with our previous measurements of $|V_{cd}|$ via $D^+\to\ell^+\nu_\ell$~\cite{bes3-Dp-muv,bes3-Dp-tauv}
and $D^{0(+)}\to \pi^{-(0)}e^+\nu_e$~\cite{bes3-D0-piev,bes3-Dp-piev}.
Conversely, using our product of  $f_{+}^\eta(0)|V_{cd}|$ and the $|V_{cd}|=0.22438\pm0.00044$ given by CKMFitter~\cite{pdg2018} yields $f_{+}^\eta(0)=0.39\pm0.04_{\rm stat}\pm0.01_{\rm syst}$.
Table~\ref{table:FF} shows comparison of our BF and hadronic FF with various theoretical calculations for $D^+\to\eta\ell^+\nu_\ell$.
Our BF result disfavors the prediction in Ref.~\cite{dse:2013nof} by 2.6$\sigma$ but agrees with the other predictions~\cite{ylwu,dse:2015gdu,cheng,dse:2018nrs,rqm} within 1.5$\sigma$.
Our result for $f_{+}^\eta(0)$ agrees well with the predictions in Refs.~\cite{dse:2015gdu,dse:2018nrsnew}.
However, it clearly rules out the prediction in Ref.~\cite{dse:2012rcv}, and disfavors the
predictions in Refs.~\cite{ylwu,dse:2013nof,dse:2018nrs,rqm} by about $2.4{\text -}3.8\sigma$.
Using our $f_{+}^\eta(0)$ and the world average of $f_{+}^\pi(0)=0.6351\pm0.0081$~\cite{hflav2018}, we
determine the hadronic FF ratio to be $f_{+}^\eta(0)/f_{+}^\pi(0)=0.61\pm0.06_{\rm stat}\pm0.02_{\rm syst}$. This provides a valuable constraint
to improve the calculations of these hadronic FFs in lattice QCD.

\begin{table*}[htp]
\centering
\caption{\label{table:FF}
\small
Comparison of our BF (in $\times 10^{-4}$) and hadronic FF with various theoretical calculations for $D^+\to\eta\ell^+\nu_\ell$.
The first and second uncertainties are statistical and systematic, respectively.
Theoretical calculations listed in the table assume no gluon component for $\eta^\prime$. Numbers marked with $^*$ denote that
the predicted ${\mathcal B}_{D^+\to\eta e^+\nu_e}$ is listed due to no predictions for $D^+\to\eta\mu^+\nu_\mu$ in Refs.~\cite{dse:2015gdu,dse:2013nof}.}
\small
\begin{tabular}{lccccccccc} \hline
&This work
& LCSR\,\cite{ylwu}& LCSR\,\cite{dse:2015gdu} & LCSR\,\cite{dse:2013nof} &LFQM\,\cite{dse:2012rcv} & CLFQM\,\cite{cheng}&CCQM\,\cite{dse:2018nrs} & CCQM\,\cite{dse:2018nrsnew}& RQM\,\cite{rqm} \\ \hline
${\mathcal B}_{D^+\to \eta\mu^+\nu_\mu}$&$10.4\pm1.0\pm0.5$&$8.4^{+1.6}_{-1.4}$&$14\pm11^*$&$24.5\pm5.3^*$&...& $12\pm1$& 9.12 &...&12.1 \\
Difference\,($\sigma$) &...&1.0&0.3&2.6&...&1.1&1.1&...&1.5\\ \hline
$f_+^{\eta}(0)$                    &$0.39\pm0.04\pm0.01$&$0.56^{+0.06}_{-0.05}$&$0.43^{+0.17}_{-0.14}$&$0.55\pm0.05$ &$0.71\pm0.01$&... &$0.67\pm0.11$&$0.36\pm0.05$&0.547 \\
Difference\,($\sigma$) &...&2.5&0.3&2.5&7.5&...&2.4&0.5&3.8\\ \hline
\end{tabular}
\end{table*}

In summary, the SL decay $D^+\to \eta\mu^+\nu_\mu$ has been observed
by analyzing 2.93 fb$^{-1}$ of data collected at $\sqrt{s}=3.773$ GeV. The absolute BF of this decay is determined
for the first time to be ${\mathcal B}_{D^+\to \eta
  \mu^+\nu_\mu}=(10.4\pm1.0_{\rm stat}\pm0.5_{\rm syst})\times
10^{-4}$.  Using the world averaged value of ${\mathcal B}_{D^+\to\eta
  e^+\nu_e}=(11.4\pm1.0)\times 10^{-4}$ gives the BF ratio
$R_{\mu/e}= {\mathcal B}_{D^+\to\eta
  \mu^+\nu_e}/{\mathcal B}_{D^+\to\eta
  e^+\nu_e} = 0.91\pm0.13$, where
  the uncertainty is the sum in quadrature of the statistical and systematic errors,
but dominated by the statistical error. This result agrees with the SM predictions
(0.97-1.00)~\cite{ylwu,cheng,dse:2018nrsnew}, thereby implying no LFU
violation within current sensitivity.
The obtained BF can be used to determine the $\eta$-$\eta'$
mixing angle once ${\mathcal B}_{D^+\to\eta^\prime \mu^+\nu_e}$ is measured with large data samples~\cite{bes3-white-paper,belle2-white-paper} in the near future.
By analyzing the dynamics of $D^+\to \eta\mu^+\nu_\mu$ decay, the product of $f_{+}^\eta(0)|V_{cd}|$
has been determined to be $0.087\pm0.008_{\rm stat}\pm0.002_{\rm syst}$. Combining necessary inputs, we have obtained $f_+^{\eta}(0)=0.39\pm0.04_{\rm stat}\pm0.01_{\rm syst}$ and $|V_{cd}|=0.242\pm0.022_{\rm stat}\pm0.006_{\rm syst}\pm0.033_{\rm theory}$. The obtained $f_+^{\eta}(0)$ provides important data to test
various theoretical calculations, while the obtained $|V_{cd}|$ is valuable for the CKM matrix unitarity test.

The BESIII collaboration thanks the staff of BEPCII and the IHEP computing center for their strong support. This work is supported in part by National Key Basic Research Program of China under Contract No. 2015CB856700; National Natural Science Foundation of China (NSFC) under Contracts Nos.~11775230, 11475123, 11625523, 11635010, 11735014, 11822506, 11835012, 11935015, 11935016, 11935018, 11961141012; the Chinese Academy of Sciences (CAS) Large-Scale Scientific Facility Program; Joint Large-Scale Scientific Facility Funds of the NSFC and CAS under Contracts Nos.~U1532101, U1932102, U1732263, U1832207; CAS Key Research Program of Frontier Sciences under Contracts Nos. QYZDJ-SSW-SLH003, QYZDJ-SSW-SLH040; 100 Talents Program of CAS; INPAC and Shanghai Key Laboratory for Particle Physics and Cosmology; ERC under Contract No. 758462; German Research Foundation DFG under Contracts Nos. Collaborative Research Center CRC 1044, FOR 2359; Istituto Nazionale di Fisica Nucleare, Italy; Ministry of Development of Turkey under Contract No. DPT2006K-120470; National Science and Technology fund; STFC (United Kingdom); The Knut and Alice Wallenberg Foundation (Sweden) under Contract No. 2016.0157; The Royal Society, UK under Contracts Nos. DH140054, DH160214; The Swedish Research Council; U. S. Department of Energy under Contracts Nos. DE-FG02-05ER41374, DE-SC-0012069.

\clearpage
\appendix
\onecolumngrid
\section*{Supplemental material}
\setcounter{table}{0}
\setcounter{figure}{0}

Table~\ref{tab:singletagN} shows the
tag dependent ST yields in data, ST efficiencies, DT efficiencies, and signal efficiencies of $D^{+}\to \eta\mu^+\nu_\mu$.
Table~\ref{tab:tab2} shows the efficiency matrix of $D^{+}\to \eta\mu^+\nu_\mu$ across different $q^2$ intervals averaged over six tag modes.

\begin{table}[htp]
\centering
\caption{\small
Tag dependent ST yields in data, ST efficiencies,
DT efficiencies and signal efficiencies of $D^{+}\to \eta\mu^+\nu_\mu$.
The efficiencies do not include the BFs of $K^0_S$, $\pi^0$, and $\eta$.
The uncertainties are statistical only.
}\label{tab:singletagN}
\begin{tabular}{lcccc}
  \hline\hline
  ST   mode & $N_{{\rm ST}}$ & $\epsilon_{{\rm ST}}$ (\%) &   $\epsilon_{{\rm DT}}$ (\%)&
  $\epsilon_{\eta\mu\nu}=\epsilon_{{\rm DT}}/\epsilon_{{\rm ST}}$  (\%)\\  \hline
$K^+\pi^-\pi^-$        & $782669\pm\hspace{0.15cm}990$                 & $50.57\pm0.06$ &$19.00\pm0.09$                &   $37.23\pm0.18$\\
$K^+\pi^-\pi^-\pi^0$   & $251008\pm1135$                               & $26.72\pm0.09$ &$\hspace{0.15cm}9.71\pm0.08$  &   $36.02\pm0.33$\\
$K^0_S\pi^-$           & $\hspace{0.15cm}91345\pm\hspace{0.15cm}320$   & $50.39\pm0.17$ &$20.56\pm0.11$                &   $40.44\pm0.25$\\
$K^0_S\pi^-\pi^0$      & $215364\pm1238$                               & $27.25\pm0.07$ &$11.41\pm0.12$                &   $41.49\pm0.45$\\
$K^0_S\pi^+\pi^-\pi^-$ & $113054\pm\hspace{0.15cm}889$                 & $28.29\pm0.12$ &$10.15\pm0.18$                &   $35.54\pm0.64$\\
$K^+K^-\pi^-$          & $\hspace{0.15cm}69034\pm\hspace{0.15cm}460$   & $40.87\pm0.24$ &$13.66\pm0.10$                &   $33.13\pm0.32$\\
  \hline\hline
\end{tabular}
\end{table}

\begin{table*}[htp]
\centering
\caption{
Efficiency matrix (\%) of $D^{+}\to \eta\mu^+\nu_\mu$ across different $q^2$ intervals.
}\label{tab:tab2}
\begin{tabular}{lccccc}
\hline
\hline
$\epsilon_{ij}$&    1&    2&    3&    4&   5 \\\hline
1 &34.99& 1.10& 0.01& 0.00& 0.00\\
2 &1.19 &34.40& 1.66& 0.00& 0.00\\
3 &0.03 &1.53 &34.78& 2.12& 0.01\\
4 &0.02 &0.04 &1.67 &34.49& 1.58\\
5 &0.02 &0.04 &0.07 &1.86 &36.53\\
 \hline
\end{tabular}
\end{table*}

\end{document}